\documentclass[preprint,eqsecnum,aps]{revtex4}
\newtheorem{definition}{Definition}

\begin{document}
%\preprint{PHYSICS LETTERS A\hspace{2.40in}}
%\twocolumn[\hsize\textwidth\columnwidth\hsize\csname@twocolumnfalse%
%\endcsname
%\draft
\title{Bell's Inequality, Quantum Measurement and Einstein
Realism:
\\ A Unified Perspective}
%\author{John V. Corbett*}
\author{John V. Corbett}
\email[]{jvc@ics.mq.edu.au}

\address{Department of Mathematics,
Macquarie University, N.S.W. 2109, Australia}
%\author{Dipankar Home*}
\author{Dipankar Home}
\email[]{dhom@bosemain.boseinst.ac.in}

\address{Department of Physics, Bose Institute, Calcutta
700 009, India}
\date{\today}
%\date{\today}% It is always \today, today, but you may
%specify any date with \date.

\begin{abstract}
The logical foundations of Bell's inequality are
reexamined. We argue that the form of the reality condition
that underpins Bell's inequality comes from the requirement
of solving the quantum measurement problem.  Hence any
violation of Bell's inequality necessarily implies
nonlocality because of the measurement problem. The
differences in the implications of deterministic and
stochastic formulations of Bell's inequality are
highlighted. The reality condition used in Bell's
inequality is shown to be a generalisation of Einstein's
later form of realism.
%Valid PACS numbers may be entered using the
%\verb+\pacs{#1}+ command.
\end{abstract}

%\pacs{Valid PACS appear here}%
\maketitle

%\pacs{PACS number(s): 03.65.Bz, 42.50.Wm}

%]

%\narrowtext
%\twocolumn

% body of paper here
%\section{Introduction}

%\vspace{.1in}
%\vspace{.5in}
%\footnoterule
%\vspace{-.2in}
%\footnotesize
%\begin{flushleft}
%*Email address: jvc@ics.mq.edu.au
%
%$^\dagger$Email address: dhom@boseinst.ernet.in
%\end{flushleft}
%\normalsize

\section{Introduction}

It is well known that Bell's inequality (BI) is violated in
quantum mechanics (QM) when the two spatially separated
systems are in an entangled or nonfactorisable state\cite{1}.
Similarly, two systems in an entangled state underlie the
measurement problem, known as "the most important puzzle in
the interpretation of quantum mechanics " \cite{2}. This is
the problem of associating an element of "reality", such as
a "complete state", with a definite outcome of an individual
measurement.  The significant relationship between BI and the
measurement problem becomes apparent when it is noted that
the derivations of BI are based upon conditions of "local
realism'' involving the concept of a "complete" or "real"
state. 
 
In section 2 of this paper we argue that the "realism"
problem of measurement provides a more
physical motivation for adopting a "complete state"
formulation of "realism"  than has been given previously. 
Such a formulation of reality when coupled with a locality
condition leads to BI, this is discussed in sections III and
IV.

Bell's original argument justifying the reality
condition that he used was derived from the Einstein-Podolsky-Rosen
(EPR) argument\cite{3} expressed in terms of Bohm's example
(EPRB). It assumed perfect quantum mechanical correlations
between the outcomes in the two wings in addition to the
reality of the individual outcomes. This assumption is not
needed in the Clauser, Horne, Shimony and Holt derivation
\cite{5} nor in the argument given here. As emphasised by
Shimony\cite{5}, in order that the empirical validity of
local realism can be tested independently of QM, no quantum
mechanical outcomes that arise from the use of the EPRB
state, such as the input of perfect correlations, should be
used to derive BI. Bell had used this input to justify the
existence of the functions A(a, $\lambda$), B(b,
$\lambda$) by following the EPR argument. This is logically
permissible for showing an inconsistency between QM and
local realism but not for empirically refuting local realism
by testing the validity of BI. A further point against the
use of perfect correlations is that the input, viz. the
equality of magnitudes of  A (a, $\lambda$) and B(a,
$\lambda$) with opposite signs, cannot  be empirically
verified as an exact equality because of the inevitable
imprecision involved in taking measurements along perfectly
opposed directions.

However, any derivation of BI which avoids the input of
perfect correlations cannot get off the ground unless it
assumes a more complete state specification (eg by a variable
$\lambda$) than the wave function $\psi$ of QM. In the recent review
by Shimony \cite{5}, this point is stressed : "It should be
emphasized that the price of abstaining from a quantum
mechanical assumption is that the definiteness of the
functions A and B must be 'postulated', rather than
'derived' as in the argument of EPR, which Bell was
following".

In view of the importance of BI in quantum nonlocality
discussions, its use as a measure of entanglement, and its
role in quantum communications it is important to have a
logical foundation of BI that is as cogent as
possible.  We think that the following form of argument
provides this.    
  
We also clarify the differences between deterministic and
stochastic formulations of BI through the nature of
nonlocality implied by the QM violation of BI. Finally, 
in section V, we point out the difference between the
forms of realism used in the EPR argument and that used in
Einstein's later version, which was given most concisely in a
letter to Besso.

We begin by stating the crux of the quantum measurement
problem and sharpen the usual discussion by emphasising the 
gap between what is \emph{required} and what is
\emph{obtained} in the standard theory concerning the nature
of  the final ensemble of states. Then the measurement
problem is  related to the form of "realism" which is
required to address it.
  
\section{The Quantum Measurement Problem}

The measurement problem comes from the well known fact that
quantum mechanics predicts only a probability distribution
of the values that would be obtained by measuring a physical
quantity on an ensemble of systems which are all prepared
identically in a state represented by a wave function.
Hence QM predictions can only be verified by identical
measurements performed on a homogeneous ensemble of systems,
all corresponding to a common wave function. After such
measurements have been performed, the final ensemble,
comprising systems and apparatuses, needs to be such that
probabilities (relative frequencies) of various outcomes 
can be determined. However, these can only be determined if
each outcome is observationally \emph{distinct} from the
others. This requires that the post-measurement ensemble be
heterogeneous, that is, the post-measurement ensemble  is a
\emph{mixture} of \emph{distinguishable apparatus states}
with distinct apparatus states corresponding to distinct
outcomes. The acuteness of this requirement is further
emphasized by the following argument.

The very concept of \emph{objective definiteness} of a
measurement outcome means that it is \emph{inter-subjective}
and it occurs \emph{independently} of whether or not it is
observed.  In other words, the actualisation of an outcome
takes place due to an interaction between a system and an
apparatus, with \emph{no} necessary participation by an 
external observer. Once actualised or recorded, an outcome
is  ``out there'' so that it can be inspected at \emph{any}
later  instant by \emph{any} observer. The actualisation of
an  individual outcome is therefore associated with an 
``objectively real'' change in the apparatus state that is 
induced by a measurement process.  However, unless the final 
apparatus states are \emph{distinguishable}, they cannot 
correspond to the actualisation of \emph{different}
outcomes. In other words, the definiteness of an individual
outcome  requires that any particular apparatus state is
\emph{distinguishable} from all other apparatus states. 
Hence  the final ensemble needs to be \emph{heterogeneous} 
in an \emph{objectively real} sense. It is precisely this
requirement  that is \emph{not} satisfied by standard QM
description of  a measurement process.

If both system and apparatus are described quantum mechanically
and the time evolution due to a measurement interaction is
governed by a Schroedinger equation, the final composite state
is inevitably a \emph{pure entangled} state. The following is a generic feature
of all QM models of measurement. Suppose a
system is initially in a state given by the wave function
$\psi$ ($\psi=a\psi_{1}+b\psi_{2}$), that is a superposition
of, say, two eigenstates $\psi_{1}$ and $\psi_{2}$ of a
dynamical variable which is being measured. Then after an
interaction with a suitable measuring device, the final
system-apparatus state is given by the wave function
\begin{equation}
\label{1}
\Psi=a\psi_{1}\phi_{1}+b\psi_{2}\phi_{2},
\end{equation}
where $\phi_{1}$ and $\phi_{2}$ are states of the apparatus
which can be distinguished by some macroscopic means. Note
that equation(\ref{1}) implies that no
element of the system-apparatus ensemble is assigned a
separate state because all elements of the ensemble have a
common wave function. Hence in the standard formulation of
QM the ensemble is \emph{homogeneous}, implying that its
members are indistinguishable. This is clearly
\emph{inconsistent} with the fact that in any one run of a
measurement, a definite value of the apparatus
variable is obtained which corresponds to either $\phi_{1}$
or $\phi_{2}$. The question of \emph{how} to resolve this 
inconsistency given a wave function of the form (\ref{1}) 
is the quantum measurement problem \cite{6}. Unless a 
resolution is obtained, it is a logical \emph{non sequitur}
to ascribe physical significance to the computed statistical 
frequencies or probabilities of various outcomes because the 
occurrence and the definiteness of an individual outcome 
is not ensured within the standard framework of QM.

When the Schroedinger equation is kept
\emph{unmodified}, there have been two general approaches in
the literature to the problem: 
(a) those that use the idea of decoherence being induced by
an interaction of the apparatus with the environment, 
(b) those in which the state of an individual member of the
ensemble is more completely specified.

We shall argue below that decoherence alone is not sufficient
to ensure that in a single run the apparatus is left in a
state characterised by a definite outcome. In other words,
the decoherence approach does not actually produce a
heterogeneous post-measurement ensemble of the type that is
necessary to \emph{solve} the measurement problem. 
Therefore, provided one does not assume that there is a
special role for conscious awareness by which outcomes are
perceived, the remaining  option for tackling the
measurement problem while the quantum dynamics is
\emph{unmodified} is (b). Then solving
the measurement problem gives a justification of the
postulate that there is a more complete state
specification.    

Let us now examine the essentials of the decoherence
approach. A measuring apparatus is categorised as a system
whose interaction with its environment (modelled quantum
mechanically) is necessarily \emph{nonnegligible} in 
contrast to the measured system whose interaction with the
environment can be neglected. Therefore a measurement
process is described quantum mechanically by a tripartite
entanglement involving system, apparatus and environment
(environment-induced decoherence models \cite{7}). Then due
to orthogonality between the states of the environment which
are coupled to system-apparatus states, the reduced density
operator of the system-apparatus subsystem is diagonal
(i.e., an \emph{effective decoherence} occurs between
system-apparatus states). This is then taken to imply that 
a pure entangled state at the end of a measurement behaves
\emph{like} a mixture of system-apparatus states
(this approach to the measurement problem was
first suggested by Heisenberg \cite{8}).

For instance, if one includes the environment in the
entanglement described by (\ref{1}), then by tracing over
the environment states, the reduced density operator
$\rho_{SA}$ pertaining to system and apparatus states is
given by
\begin{equation}
\label{2}
\rho_{SA}=|a|^{2}\left|\psi_{1}\rangle\langle\psi_{1}\right|
\otimes\left|\phi_{1}\rangle\langle\phi_{1}\right|+|b|^{2}
\left|\psi_{2}\rangle\langle\psi_{2}\right|
\otimes\left|\phi_{2}\rangle\langle\phi_{2}\right|.
\end{equation}
It is important to note that although $\rho_{SA}$ is 
diagonal it is a sum of two terms each of which describes a 
different individual outcome after a measurement. Thus the 
fact that the reduced density operator of the joint
system-apparatus system is diagonal (after tracing over
the environment states) implies only the
\emph{non-observability} of interference between apparatus
states corresponding to various outcomes (these correspond
to different diagonal elements of the reduced density
operator). The vanishing of the off-diagonal elements of
the reduced density operator is merely a  statement about
the \emph{coherence} property of the system-apparatus
states. There is no ingredient in the formalism that ensures
that in a single run  of a measurement an individual
apparatus is left in a state  characterised by a definite
outcome. The question of \emph{how} one particular diagonal
element of the reduced  density operator is picked to be the
outcome of a single measurement is not addressed.

Moreover, by simply rewriting the tripartite entangled state 
of system, apparatus and environment in a different basis,
it is possible to interpret the \emph{same} reduced density
operator of system-apparatus as corresponding to a 
different probabilistic mixture of superpositions of
system-apparatus states resulting from the \emph{same}
measurement interaction.  Thus the decoherence approach has
an additional ambiguity in failing to uniquely specify how a
particular mixture, out of the myriad of possible mixtures,
emerges for a given measurement interaction. Therefore there
is a problem of uniqueness of the interpretation of the
outcomes of a given measurement interaction even after the
environment states have been traced over because a relevant 
eigenbasis for a given measurement interaction cannot be 
uniquely identified. This ambiguity cannot be resolved unless 
a particular basis is chosen or \emph{preferred} by some 
criterion for determining the set of outcomes obtained in 
a specific measurement. This point has been stressed, 
for example, by Penrose in his debate with Hawking \cite{18}.

To sum up, although by considering an interaction with the
environment it is possible to explain why the joint
system-apparatus pure state behaves with respect to the
interference properties \emph{as if} it were a mixture  of
states, the crucial requirement that the system-apparatus
ensemble after the measurements be \emph{heterogeneous} is
not fulfilled.

There are a number of models/interpretations of quantum 
mechanics: the statistical interpretation\cite{9}, many
worlds\cite{10}, consistent histories\cite{11} and their
many variants ( see for example\cite{12}), all of which
basically invoke the idea of decoherence, either explicitly
or implicitly, in tackling the measurement problem. Hence
the criticism above applies to all of these approaches\cite{13}.

In approaches of type (b) given an ensemble that is
described by a wave function the state of an
\emph{individual} member of the ensemble is specified in a
more complete way than that provided by the wave
function. Details of \emph{how} such a specification can be
provided (whether by introducing suitable ontological
variables supplementing a wave function 
\emph{a la} Bohm's model\cite{14}, or by using open subsets
of states close to the wave function's pure state\cite{19})
are \emph{not} relevant to our subsequent discussion. It is
enough to consider the possibility of a specification in
terms of what  may be called a \emph{complete}
\emph{state} which is postulated to provide a
complete description of the state of an individual
system independent of any measurement. Then the 
heterogeneity of the post-measurement ensemble emerges
because the measurement process is treated on the same
footing as any other physical interaction and no cut
is made between a system and an apparatus.  Different
outcomes will be  characterised by different labellings  of
the complete state. It is this idea of a complete state that
underlies what we call Bell realism.

In general, this form of realism requires that a complete
state description of an individual observed system
or an apparatus is one that is \emph{sufficiently complete}
to causally connect to an individual outcome. The
argument starts by noting that a definite outcome must be
associated with a postmeasurement complete state of the
apparatus. This postmeasurement apparatus state must then be
coupled to a postmeasurement complete state of the observed
system. These postmeasurement states and their coupling
should emerge from premeasurement complete states of the
apparatus and system through the measurement process.
Therefore the definite outcome of a measurement is causally
related to the complete state that the observed system was
in before the measurement. In essence, the reality of the
complete state of the apparatus after the measurement
requires that the observed system was in a complete state
before the measurement.   

\section{Bell Realism}

The connection between the measurement problem and Bell
realism may now be summed up as follows. Since
outcomes are "objectively real" and are inferred from
differences between pre- and post-measurement apparatus
recordings, the complete state of any individual apparatus
must be "objectively real' \emph{both} at the pre- and
post-measurement level. If \emph{both} the measured system
and apparatus are treated on an \emph{equal footing} in
terms of complete states and by viewing a measurement
interaction as a physical interaction, "objective reality"
must be ascribed to the complete states of the measured
system, both before and after the measurement. Then, in such
a treatment, the outcome of measuring a dynamical variable is
causally related to a complete state of the pre-measured
system. A complete state of the system thus gives a
description of an individual system which is more complete
than that given by a QM wavefunction in the sense that it
causally corresponds to an individual outcome of a
measurement.  

The physical import of this inference comes from the
following \emph{counterfactual} implication of ascribing
"objective reality" to the correspondence between an outcome
and the complete state of the pre-measured system:

\begin{definition}Bell Realism:
If an ensemble of systems is prepared
\emph{identically} in a complete state, a particular
outcome of a measured variable will  be
obtained \emph{either} with certainty, \emph{or} with a
specific probability.
\end {definition}
The above statement is a condition of "realism", henceforth
designated as BR, which we call Bell realism. It is
\emph{deterministic} or \emph{stochastic}, depending upon
whether the correspondence between outcomes and 
complete states is respectively \emph{one-to-one} or
\emph{probabilistic}. There is no locality condition 
involved in its formulation. 

To put it mathematically, Bell realism means "objective
definiteness" of the functions of the parameters
characterizing pre-measurement complete states; such
functions correspond to definite individual outcomes of
measurements. It is precisely this form of the reality
condition that is used in conjunction with a locality
condition, LC, to derive BI. The measurement problem therefore
justifies the use of complete states that is  basic to this
type of derivation of BI. In other words, measurements are
characterised as a class of physical process whose complete
and self-consistent treatment requires that the QM framework
be augmented by introducing the notion of complete states at
both the  micro and macro levels.

\section{Bell's inequality, Deterministic and Stochastic
forms}

We shall first consider the \emph{deterministic} form of BR
in order to make our key point about the nature of nonlocality
implied by the QM violation of BI. Subsequently, we shall point 
out implications of our approach for the stochastic form of BR.

The relevant LC contingent on the deterministic form of BR 
may be stated as follows:
\begin{definition}Locality Condition:
The one-to-one correspondence between an outcome and the
complete state of a measured system is \emph{independent} of
what measurement is performed on its spatially separated
partner with which it may have interacted in the past but is
presently noninteracting.
\end{definition}

The conjunction of BR and LC implies that for any individual
system prepared in a putative complete state and for a given
dynamical variable, there is a definite result one ``would''
get if that variable is measured, irrespective of what is
measured on a distant system. We shall now briefly
recapitulate how this leads to BI.

Let \( A_{1}, A_{2} \) and \( B_{1}, B_{2} \) be two pairs of
dynamical variables associated with a pair of systems S(A) and
S(B). The systems S(A) and S(B) have been prepared in an initial
complete joint state, FS, and are now widely separated and
non-interacting. We assume that each of the four quantities, 
$A_{1}$,$A_{2}$, $B_{1}$ and $B_{2}$ can take only two values, 
$+1$ or $-1$. Since all the members of the ensemble of joint 
systems are assumed to have been prepared in a particular FS, 
then BR implies that each of the four quantities will give a
particular value when it is measured on any member of the 
ensemble. That is, the  measured value of of each of the 
quantities $A_{1}$, $A_{2}$, $B_{1}$ and $B_{2}$ is fixed. 
Then, using LC, the outcomes of the measurements of the joint
quantities \( A_{1}B_{1} \), \( A_{1}B_{2} \), \( A_{2}B_{1} \), 
and \( A_{2}B_{2} \) can be either $+1$ or $-1$.

 For each of the 16 different combinations corresponding 
to possible choices of \( \pm 1 \) for each 
\( A_{1} \), \( A_{2} \), \(B_{1} \), \( B_{2} \) the 
following equality holds 
\begin{equation}
\label{3}
A_{1}B_{1}+A_{1}B_{2}+A_{2}B_{1}-A_{2}B_{2}=\pm 2.
\end{equation}

If, on the other hand, the members of the ensemble of joint systems
were prepared in a number of different FS's, each FS being 
compatible with the restrictions imposed by the preparation procedure,
then we would have to use BR on each sub-ensemble determined by a 
specific FS to ensure that equation (\ref{3}) holds for the results 
of measurements on the systems S(A) and S(B) for each sub-ensemble.
Then summing the results, given by equation (\ref{3}) for each FS
specified sub-ensemble, over the collection of FS specified sub-
ensembles and taking an average, we obtain 
\begin{equation}
\label{4}
\left| \left\langle A_{1}B_{1}\right\rangle +\left\langle
A_{1}B_{2}\right\rangle +\left\langle A_
{2}B_{1}\right\rangle -\left\langle A_{2}B_{2}\right\rangle
\right| \leq 2.
\end{equation}

Assuming that the randomly chosen samples of particles on which
pairs of quantities such as \( A_{1}B_{1},A_{1}B_{2},... \) are
measured are typical of the entire ensemble (the principle of
induction), the averages \( \left\langle A_{1}B_{1}\right\rangle,
\left\langle A_{1}B_{2}\right\rangle,.... \) are interpretable
as the actual measured values of these quantities.

Thus BI, given by Eq. (\ref{4}), is a testable consequence of BR
in conjunction with LC. Note that the notion of counterfactual 
definiteness that underlies this derivation of BI is not merely 
counterfactual but is crucially linked to the concept of a FS 
which defines the form of realism that has been used. 

Now, it is well known that QM predictions
for entangled states violate BI for appropriate choice of
observables. In fact, it has been proved that for \emph{any}
nonfactorisable wave function of correlated quantum systems it
is always possible to choose observables so that BI is violated
by QM predictions \cite{9}. Hence in view of the way we have
formulated the conditions BR and LC that give rise to BI and
our argument that BR is \emph{necessary} to tackle the quantum
measurement problem, it follows that the incompatibility with
LC implied by QM violation of BI means the following:

The correspondence between an outcome and the FS of a system
depends on \emph{what} measurement is done on the system's
spatially separated partner. In other words, the correspondences
between outcomes and FS cannot be specified in a mutually
independent way for the ``parts'' of a composite (entangled)
quantum system, even if the ``parts'' may be widely separated.
This implies that the very form of ``connection''
between FS of ``parts'' depends on the entangled wave function
for the state of the ``whole'' (nonreductionism) - this is what
constitutes the essence of quantum nonlocality entailed by Bell's
theorem. \emph{How} such a ``connection'' is analysed or whether
it actually implies a ``mechanical'' influence across ordinary
space-time depends on the kind of model one uses for
characterising FS.

We now turn to consider the \emph{stochastic} form of BR. It is
well known that BI then follows \cite{10} if, in conjunction
with BR, one uses the following assumptions concerning the
relevant probabilities occuring in the correspondences
between outcomes and FS:

(a) The probabilities are \emph{bounded} between 0 and 1.

(b) The joint probability of outcomes for measurements on
spatially separated systems is \emph{factorisable} with respect
to their FS. This means that the joint probability is
expressible as a product of individual probabilities of the
respective outcomes pertaining to the systems prepared in the
respective FS (traditionally, in such derivations, a FS is
labelled by the parameter \( \lambda  \) - the so-called
``hidden variable''; the factorisability is then assumed at
the level of ``hidden variables'').

(c) A specific probability associated with the correspondence
between an outcome and the FS of a measured system is
\emph{independent} of \emph{what} measurement is performed on
its spatially separated partner.

Now, we recall that in the formulation of BR, probabilistic
correspondence between an outcome and a FS physically means the
following. If a system is repeatedly prepared in the \emph{same}
FS and the \emph{same} variable is measured, a particular
outcome will be obtained with a \emph{fixed} relative
frequency. This is \emph{necessarily} bounded between 0 and 1.
The assumption (a) is thus justified. Arguments interpreting
quantum incompatibility with BI as requiring the use of
negative/complex probabilities \cite{11} in stochastic
``hidden variable'' models are therefore unacceptable.

If BI is viewed as a consequence of stochastic BR, the quantum
violation of BI may be attributed to the invalidity of either
assumption (b) or assumption (c). Note that assumptions (b)
and (c) are mutually independent. While assumption (c) is
clearly a locality condition, assumption (b) expresses what
may be called the ``separability'' condition for the
probabilities of outcomes pertaining to the respective FS.
Violation of
assumption (b) would mean that for quantum entangled states
that are incompatible with BI, probabilities are
\emph{nonfactorisable} (nonseparable) at the level of FS.
Thus, quantum violation of BI in the stochastic case implies
either nonlocality or nonseparability. However, such a
distinction cannot be made in the deterministic case.

\section{Relation to Einstein Realism}

At this stage it is instructive to compare Bell realism with
the form of realism used in the EPR argument \cite{3}.
We recall that the EPR argument is based on associating an
element of physical reality with a physical quantity,
\emph{provided} its measured value can be predicted with
certainty ``without in any way disturbing'' the system in
question. Evidently, the very formulation of this criterion
of realism (EPRR) is contingent upon a locality condition.
Therefore it rules out, \emph{by fiat}, the possibility of a
nonlocal realist explanation of quantum phenomena, in
particular, of the perfect correlations in the EPR example.
However, a counterexample is provided by the Bohmian causal
realist model of QM which is manifestly nonlocal \cite{14}.

Einstein's own version of the EPR argument, which can be 
found in his letter to Besso in 1952 \cite{4}, rests on a
criterion of realism, Einstein realism or ER, that is
\emph{independent} of the locality condition. In it Einstein
assumes a \emph{one-to-one correspondence} between a wave
function and the real (complete) state of a system. In order 
to demonstrate the incompleteness of the wave function
description Einstein uses a "reductio ab adsurdum" argument
in which, as well as the one-to-one relation, a
particular form of the locality condition is assumed.

Take two spatially separated but previously interacting
systems $S_{1}$ and $S_{2}$ with a joint wave
function $\psi_{12}$. Depending on the kind of measurement 
made on $S_{1}$ and the particular measurement result, one
can subsequently ascribe a certain wave function $\psi_{2}$
to the other system $S_{2}$. Assuming the collapse
of a wave function, $\psi_{2}$ can thus take different forms
according to the kind of measurement done on $S_{1}$. Then
Einstein applied the locality condition in the form that the
complete state of $S_{2}$ should not be affected by what is
done on $S_{1}$ which is spatially separated from $S_{2}$
and no longer interacting with it. Hence he inferred that
there was no one-to-one correspondence between the complete
state of $S_{2}$ and $S_{2}$'s wave function $\psi_{2}$ which
varied depending on the measurement performed on $S_{1}$. 
This contradiction could only be avoided if the locality
condition was dropped which was unacceptable for Einstein.
Therefore he inferred that the standard formulation of QM
was incomplete.

The weakness of Einstein's argument is that within the
standard framework of QM it can be argued that although  
$\psi_{2}$ changed when different measurements were
performed on $S_{1}$, the reduced density matrix of $S_{2}$
was unaffected and hence the observable statistical
properties of $S_{2}$ were unchanged. Then the reduced
density matrix of $S_{2}$, rather than the wave function,
could be used to get a unique correspondence with the
complete state.

The realism condition BR underlying BI avoids the above
weakness of ER by \emph{generalising} ER through an objective
reality being ascribed to an individual outcome of a
measurement. As we have argued earlier, such a reality
condition is required to address the measurement problem and
this in turn implies that an individual outcome has a causal
correspondence with the complete state of a premeasured
system. Thus while ER is formulated specifically with
reference to the question as to whether a quantum wave
function describes a complete state or not , BR is
\emph{independent} of the specifics of \emph{how} a complete
state is described.

Moreover, Einstein's locality condition is modelled in terms
of a complete state being unaffected by distant measurements,
while the locality condition (LC) leading to BI
pertains to the causal correspondence between a definite
outcome and a complete state being unaffected by
distant measurements.  This form of LC together with BR
enables BI to be derived without any reference to QM or to
any specifics of the way a real state or complete state
is modelled.

\section{Concluding Remarks}
The key elements of this paper are as follows: (i) The
central point is that the postulate of realism
underlying BI can be justified in a logically cogent way on
the basis of the quantum measurement problem, without
requiring to either assume it a priori (as is assumed in all
the derivations of BI which follow the CHSH type logic) or
by following Bell's original argument using the EPR-type 
reasoning whose logical weakness we have discussed. 

Furthermore, we have pinpointed the precise problem with 
Einstein's later version of the EPR argument. We have also
clarified in what sense Bell realism can be viewed as a
generalisation of Einstein realism, thereby avoiding that
problem.

To the best of our knowledge, such a unified perspective on 
the relationship between Bell's Theorem, Measurement Problem
and Einstein Realism is lacking in the literature. This
clarification, we believe, is helpful because it increases
our understanding of the logical foundations of BI which is
now being so widely used in diverse areas.  

(ii) Though the distinction between separability and locality
has been discussed earlier, the precise distinction between
the implications of the deterministic and stochastic
versions of BI in terms of the "in-principle impossibility
of preparing a complete state of an individual system" has
not been stressed previously in the way done in this paper.

To sum up, starting from general considerations arising from
the quantum measurement problem, we have motivated the Bell
realism  required to derive BI. Thus the nonlocality
inferred from the quantum mechanical violation of BI
\emph{cannot} be avoided if one  takes on board the quantum
measurement problem. Our analysis also helps to clarify the
distinctions between Bell realism and  the forms of realism
used in the EPR argument or by  Einstein in a way that has
not been stressed in the  literature. However, Bell realism
can be either deterministic or stochastic. Accordingly, as
we have discussed above, the implications of the quantum
violations of deterministic or stochastic BI differ as
regards the \emph{nature} of quantum nonlocality. The
significance of these differences  needs to be investigated
further.
\begin{acknowledgements}
The authors thank Manoj K. Samal of the S. N. Bose National Centre for
Basic Sciences in Kolkata for helpful discussions and acknowledge the
financial support for the visit of DH to Macquarie University through
a Research Grant to JVC from Macquarie University. The research of
D.H. is supported by the Jawaharlal Nehru Fellowship.
\end{acknowledgements}

% now the references. delete or change fake bibitem. delete next
% three lines and directly read in your .bbl file if you use bibtex.

\end{document}